\documentclass[11pt]{article}   

\usepackage{graphicx}            
\usepackage{amssymb}             
\usepackage{amsmath}
\usepackage{amsthm}              
\usepackage{mathtools}
\usepackage{geometry}
\usepackage{setspace}
\usepackage{booktabs, rotating, multirow, pdflscape }
\usepackage{subfig}
\usepackage{bbm}
\usepackage{booktabs}
\usepackage{dsfont}
\usepackage[colorlinks=true,
            linkcolor=black,
            urlcolor=blue,
            citecolor=blue]{hyperref}
            
\usepackage[affil-it]{authblk}
 
\renewcommand\Affilfont{\itshape\small}
\makeatletter
\renewcommand\AB@affilsepx{ , \protect\Affilfont}
\makeatother

\usepackage{fullpage} 
\usepackage{natbib}
\bibpunct{(}{)}{;}{a}{,}{,}
\usepackage{tikz}
\usepackage{enumerate} 

\newcommand{\vtheta}{\boldsymbol{\theta}}
\newcommand{\vY}{\boldsymbol{Y}}
\newcommand{\vC}{\boldsymbol{C}}
\newcommand{\vX}{\boldsymbol{X}}
\newcommand{\vZ}{\boldsymbol{Z}}
\newcommand{\vD}{\boldsymbol{D}}
\newcommand{\vYobs}{\boldsymbol{Y}^{obs}}

\newcommand{\vZobs}{\boldsymbol{Z}^{obs}}
\newcommand{\vDobs}{\boldsymbol{D}^{obs}}

\newcommand{\prob}{\text{p}}

\usepackage{color}

\title{Posterior Predictive $p$-values with Fisher Randomization Tests \\in Noncompliance Settings:\\
 Test Statistics vs Discrepancy Variables}
 \author[1]{Laura Forastiere } 
\author[1]{Fabrizia Mealli} 
\affil[1]{University of Florence}
\author[2]{Luke Miratrix} 
\affil[2]{Harvard University}

\date{}

\begin{document}

\maketitle


\begin{abstract}
In randomized experiments with noncompliance, tests may focus on compliers rather than on the overall sample.  Rubin (1998) put forth such a method, and argued that  testing for the complier average causal effect and averaging permutation based $p$-values over the posterior distribution of the compliance status could increase power, as compared to general intent-to-treat tests. The general scheme is to repeatedly do a two-step process of imputing missing compliance statuses and conducting a permutation test with the completed data. In this paper, we explore this idea further, comparing the use of discrepancy measures, which depend on unknown but imputed parameters, to classical test statistics and exploring different approaches for imputing the unknown compliance statuses. We also examine consequences of model misspecification in the imputation step, and discuss to what extent this additional modeling undercuts the permutation test's model independence.  We find that, especially for discrepancy measures, modeling choices can impact both power and validity.  In particular, imputing missing compliance statuses assuming the null can radically reduce power, but not doing so can jeopardize validity.  Fortunately, covariates predictive of compliance status can mitigate these results.  Finally, we compare this overall approach to Bayesian model-based tests, that is tests that are directly derived from posterior credible intervals, under both correct and incorrect model specification.  We find that adding the permutation step in an otherwise Bayesian approach improves robustness to model specification without substantial loss of power.

\end{abstract}

\section{Introduction}
Noncompliance refers to the situation when the actual treatment received does not perfectly correspond to treatment assigned in a randomized experiment.
With noncompliance, a simple intent-to-treat analysis, which compares outcomes based on the assignment, can fail to estimate the effect of the treatment itself.
The usual approach in these circumstances is to use instrumental variables (IV)\citep{Imbens:Angrist:1994, Angrist:1996}, which can be thought of as a special case of principal stratification \citep{Frangakis:Rubin:2002}, where the units are partitioned into subpopulations defined by the compliance behavior, and where the focus is then on the effect of the treatment among compliers, i.e., those who would always comply to the treatment assigned irrespective of the arm. 
This subpopulation effect is often referred to as the complier average causal effect (CACE). 
In an IV analysis, identification of such effect hinges on the absence of defiers (monotonicity assumption) and the lack of any effect of the assignment for noncompliers (the exclusion restriction assumption). 
Under these assumptions, a zero intent-to-treat effect (ITT) is a necessary and sufficient condition for CACE being zero as well. 
Therefore, a hypothesis test for ITT will be a valid test to assess a zero treatment effect for compliers. 
Nevertheless, ITT tests ignore all compliance information. 
This is where \cite{Rubin:1998} attempted to obtain more power by focusing on testing procedures that incorporate compliance information.
This was motivated by the observation that, taking the presence of noncompliance into account, rather than just estimating an overall average effect, would exploit the information of the data to a larger extent.
On the other hand, such a test seems challenging to construct because the compliance status is typically not known for all units.


Rubin's proposal was to average $p$-values over the posterior predictive distribution of the vector of compliance statuses.
Under an MCMC perspective this translates into including an imputation or data augmentation step, where each unit's compliance status is imputed from its posterior predictive distribution and then a $p$-value for the complier effect is generated from the complete data within each MCMC iteration.
This Bayesian approach to $p$-values has its roots in the posterior predictive model-checking method \citep{Guttman:1967, Rubin:1981,Rubin:1984}, a popular model-checking tool that can account for nuisance parameters by comparing the observed data to synthetic data drawn from the posterior predictive distribution of a hypothesized model marginalized over the nuisance parameters. 
\citet{Meng:1994} and \citet{Gelman:1996} extended posterior predictive checks by replacing classical test statistics with discrepancy variables, that is, variables that depend on the nuisance parameters, and this extension can also be applied to testing. 
Finally, \citet{Rubin:1998} applied the idea of Bayesian posterior $p$-values to Fisher randomization tests \citep{Fisher:1925, Fisher:1926, Fisher:1935} for experimental designs in the presence of noncompliance. 
Fisher randomization tests (FRTs), or permutation tests, are nonparametric tests that can be used to test a null hypothesis on the outcome distribution of a randomized experiment.
The reference distribution of a test statistic is derived by computing its value for each possible permutation of the assignment vector under the known assignment mechanism. 

In this paper, we explore this general idea of posterior predictive Fisher randomization tests more in depth, and conduct extensive simulation studies to show how these tests play out in practice in randomized experiments with noncompliance.
The combination of Fisher randomization tests with posterior predictive $p$-values lead to a sequence of  both a permutation and an imputation step. 
At each iteration, 
a test statistic, in its classical view, is computed from the data that would have been observed under the permuted assignment vector and the imputed compliance status vector.
\citet{Rubin:1998} proposed the use of any estimator of the complier average casual effect as classical tests statistic.
Here, we first examine the replacement of such test statistics with discrepancy measures.
In noncompliance settings, a discrepancy measure for obtaining posterior predictive FRTs would be an estimate of CACE conditional on the imputed compliance statuses. 
These measures seem promising because they can directly estimate CACE from the complete data.
We investigate the benefits and disadvantages of using discrepancy measures for testing as compared to test statistics.

We also  closely examine the imputation step.
Different methods are possible, such as imputing under the null or under the alternative.
Imposing the null seems a natural choice from a testing approach, and is also in line with posterior predictive checks, where nuisance parameters are typically drawn from the assumed model.
Imposing the null also protects the validity of the overall testing procedure.
However, this approach turns out to have some potential costs in terms of power.
We explore this, and discuss how to mitigate these costs.

The imputation step without the permutation step is nothing more than what would typically be used for the direct estimation of the posterior distribution of CACE. 
Credible intervals of this posterior distribution could themselves lead to nominal $p$-values. 
These model-based $p$-values are less computationally demanding than the posterior predictive $p$-values obtained by a Fisher randomization test within the imputation step.
We compare these two approaches, and see if the addition of the Fisher randomization test has anything extra to offer.
In particular, we compare the performance of these two testing methods under both correct and incorrect model specification.
Simulation studies suggest several ``best practices'' as well as help elucidate the reasons of why each approach can work.

Finally, we discuss how predictive covariates are of particular importance for the performance of discrepancy-based FRTs. 
Predictive covariates help alleviate model misspecification concerns which can arise in the imputation step, and help address many of the concerns found in our investigations.

The paper is organized as follows. 
In Section \ref{sec:ppc}, we give a brief review of posterior predictive checks. 
We then introduce Fisher randomization tests in Section \ref{sec:FRT} and illustrate the use of posterior predictive $p$-values to deal with the unknown compliance status in noncompliance settings. 
In Section \ref{sec:sim}, we describe the simulation studies we constructed to compare the use of discrepancy measures versus tests statistics, assess the impact of different modeling assumptions, and determine the potential benefits of  predictive covariates. 
Simulation results are shown in Section \ref{sec:res}. 
In Section \ref{sec:bb} we compare the validity of FRTs in combination with posterior predictive $p$-values to the corresponding Bayesian model-based tests. 
We discuss common patterns and what they suggest for practice, in Section \ref{sec:conc}.

\section{Posterior Predictive Checks}
\label{sec:ppc}
Classical $p$-values were extended to the Bayesian framework by \citet{Guttman:1967} and \citet{Rubin:1981,Rubin:1984}. 
The Bayesian framework is particularly appealing for investigating the compatibility of a posited model with observed data when the model has unknown nuisance parameters (composite null model).
While classical method would typically plug-in a point estimate of the parameter and rely on known reference distributions of pivotal quantities or on asymptotic results, Bayesian tests average over the posterior distribution of the unknown  parameters and use the posterior predictive distribution to simulate the reference distribution for any test statistic.


Suppose we have a realization $Y^{obs}$ of a random variable $Y$ and we posit a parametric \textit{null} model, 
$H_0: \vY \sim f(\vY\mid \vtheta), \vtheta \in \Theta_0$. The essence of model assessment lies in comparing the observed data with hypothetical replicates that could be observed under the assumed model. 
The classical approach amounts to measuring the discrepancy between the observed value of a test statistic, $T(\vYobs)$, and its reference (i.e., sampling) distribution derived under the posited model. 
A Bayesian model checking approach uses the posterior predictive distribution under the null hypothesis, $p(\vY|\vYobs, H_0)$, to derive the posterior distribution of the test statistic. Assuming a test statistic $T(\vYobs)$ where larger values contraindicate the null, the posterior predictive $p$-value based on the test statistic, $p_{B_T}$, is then
\begin{equation}
\begin{aligned}
\label{eq:pBT}
p_{B_T}=&Pr_{\vY}\left\{T(\vY)\geq T(\vYobs) \mid \vYobs, H_0\right\} \\
=&E_{\vtheta}\left\{Pr_{\vY}\left\{T(\vY)\geq T(\vYobs) \mid \vYobs, H_0, \theta \right\} \mid \vYobs, H_0 \right\} .
\end{aligned}
\end{equation}
The presence of unknown parameters has been taken into account by averaging over their posterior distribution under the null hypothesis. 
A Monte-Carlo simulation-based approach would draw $K$ values of the parameters, \{$\vtheta^k;\,\,\,  k=1,\ldots,K\}$, from their posterior distribution $\pi(\vtheta|\vYobs, H_0)$, simulate replications of the data under the conditional distribution $f(\vY\mid \vtheta^k)$ and compare the new values of the test statistic with the observed value. 
This approach follows from an equivalent expression of equation \eqref{eq:pBT}:
\begin{equation}
p_{B_T}= \int_{\vtheta \in \Theta_0}  \int_{\vY}\mathbf{1}\big[T(\vY)\geq T(\vYobs)\big]f(\vY\mid \vtheta)\pi(\theta\mid \vYobs, H_0)d\vY d\vtheta
\end{equation}
where $\mathbf{1}[\cdot]$ is the indicator function.\\

%
\bigskip
Meng (1994) and later Gelman, Meng and Stern (1996) proposed to replace classical test statistics, $T(\vY)$, with parameter-dependent statistics, $D(\vY, \vtheta)$, referred to as \textit{discrepancy} variables. 
These cannot translate to the classical framework because the parameter values are unknown.
In the Bayesian framework, however, they can be used as the posterior gives us predictions for these parameters and so both the ``observed'' discrepancy variables as well as its reference distribution can be calculated.
In particular, the posterior predictive $p$-value based on a discrepancy variable, $p_{D_T}$, is
\begin{equation}
\begin{aligned}
\label{eq:pBD}
p_{B_D}=&E_{\vtheta}\left\{Pr_{\vY}\left\{D(\vY, \vtheta)\geq D(\vYobs, \vtheta) \mid \vYobs, H_0, \vtheta \right\} \mid \vYobs, H_0 \right\}\\
=& \int_{\vtheta \in \Theta_0}  \int_{\vY}\mathbf{1}\big[ D(\vY, \vtheta)\geq D(\vYobs, \vtheta)\big]f(\vY\mid \vtheta)\pi(\theta\mid \vYobs, H_0)d\vY d\vtheta .
\end{aligned}
\end{equation}
Note how both $D(\vY, \vtheta)$ and $D(\vYobs, \vtheta)$ are random under the posterior distribution.
This approach has two advantages. First, a discrepancy variable requires smaller computational effort than a test statistic, given that the latter often involves an additional estimation of the parameters. 
Second, classical test statistics are typically computed by plugging-in an estimate of the nuisance parameter, thus they assess the `discrepancy'  between the data and the best fit of the model. Conversely, the use of a parameter-dependent statistic directly checks the `discrepancy' between the data and the overall model. 

\citet{Meng:1994} and \citet{Robins:2000} derived several results on the frequency evaluation of discrepancy $p$-values under the null. If $D(\vY, \vtheta)$ is a pivotal quantity with known distribution $\mathcal{D}_0$ under the null, then the distribution of $p$-values under the null would still be uniform and their expression would simplify to $p_{B_D}=Pr\left\{\mathcal{D}_0 \geq D(\vYobs, E[\vtheta\mid \vYobs, H_0]) \right\}$. 
On the other hand, in the more common situation where the discrepancy is not a pivotal quantity, averaging over the parameters on which the discrepancy depends leads to a distribution of $p$-values that is no longer uniform. Meng investigates the behavior of such $p$-values under the prior predictive distribution conditional on the null, i.e., $p(\vY\mid H_0)$. His main result is that, under this distribution, discrepancy $p$-values are centered around $\frac{1}{2}$, i.e., $E_{\vY}\{p_{B_D} \mid H_0\}=1/2$ and that $Pr_{\vY}\left\{p_{B_D} \leq \alpha  \mid H_0 \right\} \leq 2\alpha$. 
This means that there are cases in which $p$-values are conservative and other cases in which they are anti-conservative, but there is a bound for the Type \textrm{I} error of twice the nominal level. 
His further discussion suggests, however, that in practice we would expect error rates to rarely be this high.
As the posterior $p$-values are stochastically less variable than $U[0,1]$, we expect the tails to be lighter, and the error rates to often be conservative for low values of $\alpha$. 

Extending this work, {\citet{Robins:2000} show that discrepancy-based $p$-values can be seriously conservative even when the discrepancy has asymptotic mean 0 for all values of the nuisance parameters, whereas posterior predictive $p$-values based on test statistics are conservative whenever the asymptotic mean of the test statistic depends on the parameters.
Arguably, the conservativeness of a test is not a bad thing per se, because it means that it would not reject a true hypothesis more often than indicated by the nominal level. 
Indeed, such tests are considered valid \citep{Neyman:1934}, as the Type \textrm{I} error would be less or equal to $\alpha$. 
According to \citet{Rubin:1996a}, the typical conservativeness when using discrepancies, noted by \citet{Meng:1994} and \citet{Gelman:1996}, arises from the `extra' information carried by the imputations of $\vtheta$.
This information can be traced to both modeling and structural assumptions used to define the posterior distributions used for the imputation; a fundamental role is played by the the fact that imputations are performed under the null hypothesis. 
This argument can be connected to the one for potential conservatism of multiple imputation in \citet{Rubin:1996b}, where these informative imputations are called `superefficient'. 



%



\section{Fisher Randomization Tests using Posterior Predictive $p$-values}
\label{sec:FRT}
\citet{Fisher:1925, Fisher:1926, Fisher:1935} proposed a distribution-free technique to test a sharp null hypothesis of zero treatment effect at the unit level for randomized experiments.
\citet{Rubin:1984} then showed how these Fisher's randomization tests (FRTs) can be formally viewed as a posterior predictive check. 
The Bayesian justification is that they are based on the posterior predictive distribution of the test statistic induced by the random assignment $\vZ$.
Although Fisher never used the potential outcomes framework---a method of articulating causal effects originally proposed by
Neyman in the context of randomized experiments \citep{Neyman:1923} and then formalized and extended to observational studies by Rubin \citep{Rubin:1974, Rubin:1978}---FRTs can be phrased in terms of potential outcomes. 

Under the potential outcome framework, the potential outcomes, denoted $Y_i(1)$ and $Y_i(0)$, represent the
outcomes for individual $i$ had he received the treatment ($Z_i=1$) or control ($Z_i=0$) respectively. Let $Z^{obs}_i$ be the treatment that was actually assigned to unit $i$. 
The ``fundamental problem of causal inference'' \citep{Holland:1986} is that, for each individual, we can observe only one of these potential outcomes, i.e., $Y_i^{obs}=Y_i(1)Z^{obs}_i +Y_i(0)(1-Z^{obs}_i)$, because each unit will receive either treatment or control. 
As first formalized in \citet{Rubin:1974}, all causal effects are inherently a comparison of potential outcomes. Thus, Fisher's sharp null hypothesis $H_0$ of no treatment effect can be formalized as $Y_i(0)=Y_i(1) \, \, \forall i$ and Fisher's $p$-values can be stated as $Pr_{\vZ}\left\{T(\vY(\vZ),\vZ)\geq T(\vYobs, \vZobs) \mid \vZobs, \vYobs, H_0 \right\}$. Fisher's hypothesis is said to be sharp because it allows one to perfectly impute $Y_i(1-Z_i)$ for any value of $Z_i$ (i.e., the missing potential outcome). 

\citet{Rubin:1998} then formalized the idea of Fisher randomization-based tests using posterior predictive $p$-values (FRT-PP) in the context of noncompliance. Let $D_i$ be the actual treatment received by unit $i$; with noncompliance $D_i$ may differ from $Z_i$. The compliance type for each unit is then defined by the joint values of the treatment receipt if assigned to control, $D_i(0)$, or to treatment $D_i(1)$. Because we can never observe both $D_i(0)$ and $D_i(1)$, however, the compliance status is not generally known for all units. 
These are the unknown variables we will average over to obtain posterior predictive $p$-values. 
\cite{Rubin:1998} took this approach for the case of one-sided noncompliance, where $D_i(0)=0$ for all units.
With one-sided noncompliance, we have two compliance types: `compliers', for whom $D_i(1)=1$ and `never-takers', for whom  $D_i(1)=0$. In this case compliance statuses are unknown only for those units in the control arm. 
We also discuss one-sided noncompliance.

Let $C_i$ denote the compliance status indicator, being $0$ for a never-taker or $1$ for a complier.  The complier average causal effect (CACE) can then be written as
\begin{equation}
CACE:=E[Y_i(1)-Y_i(0) | C_i=1]
\end{equation}.
Assuming the exclusion restriction for non-compliers, i.e., $Y_i(0)\!=Y_i(1) \,\, \forall i\!\!: C_i=0$, 
 the null hypothesis we wish to test here is a null effect for compliers, i.e., $H_0: Y_i(0)=Y_i(1) \,\, \forall i\!\!: C_i=1$. It is worth noting that, in one-sided noncompliance settings under the exclusion restriction, CACE can be expressed as the ratio between the intent-to-treat effect, i.e., $ITT=E[Y_i(1)-Y_i(0)]$, and the probability of being a complier, i.e, $\pi_c=Pr(C_i=1)$. Therefore, in principle a rejection of a zero ITT would necessarily mean a rejection of a zero CACE. 
The goal of \citet{Rubin:1998} was to show that in some circumstances tests that take noncompliance into account can be more powerful than the ones based on the ITT only.  
 

Rubin proposed the use of a test statistic $T$, which depends on the observed data $O(\vZ)$ for each assignment vector $\vZ$, with $O(\vZ)=[\vY(\vZ), \vD(\vZ), \vZ]$, and not on the imputed compliance statuses. 
Test statistics can be any estimator of the complier average causal effect (CACE): posterior mean, posterior median or posterior mode as well as MLE or IV estimates. 
Regardless of the choice, the Bayesian procedure averages $p$-values over the posterior predictive distribution of the unknown compliance statuses, $\prob(\vC\mid \vYobs, \vDobs, \vZobs, H_0)$, which will in turn depend on other unknown parameters:
 \begin{equation}
\label{eq:pBT_c}
p_{B_T}=E_{\vtheta}\left\{E_{\vC}\left\{p_{B_T}(\vC) \mid \vYobs, \vDobs, \vZobs, \vtheta \right\} \mid  \vYobs, \vDobs, \vZobs, H_0 \right\} ,
\end{equation}

where
 \begin{equation*}
 \begin{aligned}
p_{B_T}(\vC)&=Pr_{\vZ}\left\{T(\vY(\vZ), \vD(\vZ), \vZ)\geq T(\vYobs, \vDobs, \vZobs) \mid  \vYobs, \vDobs, \vZobs, \vC, H_0 \right\}\\
&=Pr_{\vZ}\left\{T(\vYobs, \vC\vZ, \vZ)\geq T(\vYobs, \vDobs, \vZobs) \mid \vC \right\} .
\end{aligned}
\end{equation*}
The last expression follows from two observations: (1) under the null hypothesis $Y_i(Z_i)$ is equal to the observed outcome; and (2) $D_i(Z_i)=C_i Z_i$ thanks to the definition of the compliance status $C_i$ in one-sided noncompliance settings. 

We can also use the equivalent expressions of
\begin{equation}
\begin{aligned}
p_{B_T}= \int_{\vtheta \in \Theta} \int_{\vC}\int_{\vZ} &\bigg[\int_{\vY(\vZ)}\int_{D(\vZ)}\mathbf{1}\big[ T\big(\vY(\vZ), \vD(\vZ), \vZ\big)\geq T\big(\vYobs, \vDobs, \vZobs\big)\big]\times\\
&\quad \prob(\vY(\vZ), \vD(\vZ)\mid \vYobs,\vDobs, \vZobs, \vC,H_0)\,d\vY(\vZ)\,d\vD(\vZ)\bigg] \\
&\prob(\vZ)\prob(\vC \mid \vYobs, \vDobs, \vZobs, \vtheta)\pi(\theta\mid \vYobs, \vDobs, \vZobs, H_0)\,d\vZ \,d\vC\, d\vtheta ,
\end{aligned}
\end{equation}
or
\begin{equation}
\begin{aligned}
p_{B_T}= \int_{\vtheta \in \Theta} \int_{\vC}\int_{\vZ} &\mathbf{1}\big[ T\big(\vYobs, \vC\vZ, \vZ\big)\geq T\big(\vYobs, \vDobs, \vZobs\big)\big]\\
&\prob(\vZ)\prob(\vC \mid \vYobs, \vDobs, \vZobs, \vtheta)\pi(\theta\mid \vYobs, \vDobs, \vZobs, H_0)\,d\vZ \,d\vC\, d\vtheta .
\end{aligned}
\end{equation}


An MCMC approach proceeds, at each iteration $k$, by first drawing the parameters from their posterior distribution and then imputing the missing compliance statuses in the control arm using the posterior predictive distribution conditional on these draws. 
Then, for the permutation step, the assignment vector is permuted and the test statistic computed based on the outcome and treatment that would be observed under the new assignment vector, if the imputed compliance statuses were true and the individual treatment effect were null, i.e.,  $T^k=T\big(\vYobs, \vC\vZ, \vZ\big)$.
The $p$-value $p_{B_T}$ is then the proportion of iterations where the test statistic $T^k$ is greater than or equal to the observed statistic $T^{obs}=T\big(\vYobs, \vDobs, \vZobs \big)$.

\medskip
 
Following \citet{Meng:1994} and \citet{Gelman:1996}, we explore replacing parameter-independent test statistics with parameter-dependent discrepancy variables. 
Rubin (1998) already mentioned the possibility of using discrepancies, such as difference-in-means estimate of the effect among compliers, i.e. $\overline{Y}_{1,c}-\overline{Y}_{0,c} $, but he only used test statistics dependent on $O(\vZ)$ in his examples. 
Compliance-dependent discrepancy $p$-values can be written as:

\begin{equation}
\label{eq:pBD_c}
p_{B_D}=E_{\vtheta}\left\{E_{\vC}\left\{p_{B_D}(\vC) \mid \vYobs, \vDobs, \vZobs, \vtheta \right\} \mid  \vYobs, \vDobs, \vZobs, H_0 \right\}
\end{equation}
where
\begin{equation*}
\begin{aligned}
p_{B_D}(\vC)&=Pr_{\vZ}\left\{D(\vY(\vZ), \vC, \vZ)\geq D(\vYobs, \vC, \vZobs) \mid  \vYobs, \vDobs, \vZobs, \vC, H_0 \right\}\\
&=Pr_{\vZ}\left\{D(\vYobs, \vC, \vZ)\geq D(\vYobs, \vC, \vZobs) \mid \vC \right\} .
\end{aligned}
\end{equation*}


In the MCMC, at each iteration the imputed compliance statuses will affect both values of discrepancy variable, the one for the permuted assignment vector under the null hypothesis, $D^k=D(\vYobs, \vC, \vZ)$, and the one for the observed values, $D^{k,obs}=D(\vYobs, \vC, \vZobs)$. $p_{B_D}$ is then the proportion of iterations where $D^k\geq D^{k,obs}$.

Our purpose here is to compare the frequentist performance of FRT-PP based on test statistics $T(\vY(\vZ), \vD(\vZ), \vZ)$ or on imputed compliance-dependent discrepancy variables $D(\vY(\vZ), \vC, \vZ)$. In particular, we will use the typical IV estimator of CACE \citep{Imbens:Angrist:1994, Angrist:1996} as our test statistic:


\begin{equation}
\label{eq:T}
T(\vY(\vZ), \vD(\vZ), \vZ)=\frac{\widehat{ITT}_Y}{\hat{\pi}_c}=\frac{\overline{\vY}_{1}-\overline{\vY}_{0}}{\overline{\vD}_{1}-\overline{\vD}_{0}} ,
\end{equation}
where $\widehat{ITT}_Y$ is the method of moments estimator of the effect of the assignment on the outcome, $\hat{\pi}_c$ is the method of moments estimator of the proportion of compliers, 
\[\overline{\vY}_{1}=\frac{\sum_{i}Y_i(Z_i)Z_i}{\sum_{i}Z_i} \qquad\overline{\vY}_{0}=\frac{\sum_{i}Y_i(Z_i)(1-Z_i)}{\sum_{i}(1-Z_i)} \]
and 
\[\overline{\vD}_{1}=\frac{\sum_{i}D_i(Z_i)Z_i}{\sum_{i}Z_i} \qquad\overline{\vD}_{0}=\frac{\sum_{i}D_i(Z_i)(1-Z_i)}{\sum_{i}(1-Z_i)} . \]
Similarly, the discrepancy variable will be the method of moment estimator of the complier average causal effect if compliance status were known:
\begin{equation}
\label{eq:D}
D(\vY(\vZ), \vC, \vZ)=\overline{\vY}_{c1}-\overline{\vY}_{c0}
\end{equation}
where
\[\overline{\vY}_{c1}=\frac{\sum_{i}Y_i(Z_i)C_iZ_i}{\sum_{i}C_iZ_i} \qquad\overline{\vY}_{c0}=\frac{\sum_{i}Y_i(Z_i)C_i(1-Z_i)}{\sum_{i}C_i(1-Z_i)} \]

\medskip
Based on our previous discussion on the typical conservativeness of discrepancy-based $p$-values, we expect the use of these $p$-values for problems of noncompliance to lead to conservative tests. Simulations under the null, not shown here, confirm our hypotheses. In fact, in the hypothetical situation where compliance statuses were known for all units, the distribution of $p$-values would still look uniform. On the other hand, when compliance statuses were imputed for units assigned to control, the distribution of $p$-values seemed to be concentrated around 0.5. We tested this result in both the hypothetical case where 
compliance statuses are imputed from the correct model with known parameters $\prob(\vC\mid \vYobs, \vDobs, \vZobs; \vtheta^{\star})$ and the more realistic situation where the parameters are not known.
In the latter case, the Bayesian procedure that follows from the definition of $p$-values in \eqref{eq:pBT_c} and \eqref{eq:pBD_c} uses the posterior predictive distribution of the unknown compliance statuses, which in turn depends on the posterior distribution of the parameters:
$\prob(\vC\mid \vYobs, \vDobs, \vZobs, H_0)= \int_{\theta} \prob(\vC\mid \vYobs, \vDobs, \vZobs, \vtheta) \pi(\vtheta\mid \vYobs, \vDobs, \vZobs, H_0) d\vtheta$.
Imputation was still conducted under the correct model.
The conservativeness that was seen in this case is presumably due to the fact that the imputed replications of compliers would carry information on the null hypothesis, given that the parameters were estimated under the correct imputation model (consisting of a correct model specification, the exclusion restriction assumption, and the null hypothesis). In fact, the discrepancy measure makes use of the information provided by both the data and the model that is stored in the imputed compliance statuses. 


Most of the past literature focuses on the comparison of $p$-values based on discrepancies and classical statistics under the null hypothesis. There seems to be less work concerning power. 
We should bear in mind that type \textrm{I} error and power are usually traded off against each other: 
a test that is less likely to reject a correct hypothesis is usually also less likely to reject an incorrect one.
The purpose of this paper is to shed light on this trade-off for discrepancy-based $p$-values in noncompliance settings.
Does a reduction in the type \textrm{I} error to below $\alpha$ levels signal a loss of power? When is this trade-off more of an issue?

Power simulations in the hypothetical case with known parameters and correct (alternative) model specification showed that discrepancy measures can give a large increase in power.
However, in the more realistic situation with unknown parameters, the literature on posterior predictive $p$-values suggests to estimate the parameters under the null hypothesis. 
Conditioning on the null-hypothesis is motivated by the need for obtaining a valid test under the null. 
However, under a true alternative, imputation of compliance statuses would then be conducted under the wrong model which could result in a loss of power to detect a non-zero CACE.
The intuition is that there might be scenarios where the distributions of outcomes for compliers and never-takers under active and control treatment assignment are such that in the control arm units imputed as compliers from the posterior distribution conditional on the null hypothesis end up being the ones with outcomes close to the compliers in the treatment group, resulting in small values of the observed complier average causal effect, i.e., $D(\vYobs, \vDobs, \vZobs)$. 
This potential problem motivated us to investigate an imputation of the compliance statuses from the posterior distribution $\prob(\vC\mid \vYobs, \vDobs, \vZobs)$ without imposition of the null as an alternative.
This is akin to a ``plug-in'' style approach in classical testing.
However, due to the same trade-off mentioned earlier, this relaxation could lead to an increase in type \textrm{I} error, possibly giving an invalid test. 
We will also investigate how the use of observed covariates, when available, can improve the imputation and thus the performance of $p$-values under either approach. 


Given these tensions, we conducted a simulation study in order to assess whether these trade-offs indeed occur, and in order to provide clear recommendations on what is best procedure. 
In particular, we wish to compare the overall accuracy of discrepancy-based and statistic-based FRT-PP under different compliance imputation models.

\section{Simulation Study}
\label{sec:sim}
Our simulation study is designed to assess the performance of different types of randomization-based $p$-values in testing the sharp null hypothesis of no treatment effect for compliers, i.e., $H_0: Y_i(0)=Y_i(1) \,\,\forall i: C_i=1$, under the exclusion restriction assumption, i.e., $Y_i(0)=Y_i(1) \forall i: C_i=0$. 
Specifically, we compared $p$-values from  \eqref{eq:pBT_c} using the test statistic in \eqref{eq:T} and $p$-values from \eqref{eq:pBD_c} based on the discrepancy variable in \eqref{eq:D}. 
We also used four different methods for imputing the compliance status in the control arm:  
\begin{enumerate}[$$]
\item Imputation method 1: impute imposing the null hypothesis without including covariates, that is, use the posterior distribution 
$\prob(\vC\mid \vYobs, \vDobs, \vZobs, H_0)$
\item Imputation method 2: impute without imposing the null hypothesis and without including covariates, that is, use the posterior distribution 
$\prob(\vC\mid \vYobs, \vDobs, \vZobs)$
\item Imputation method 3: impute imposing the null hypothesis and including covariates, that is, use the posterior distribution 
 $\prob(\vC\mid \vYobs, \vDobs, \vZobs, H_0, \vX)$
\item Imputation method 4: impute without imposing the null hypothesis and including covariates, that is, use the posterior distribution 
 $\prob(\vC\mid \vYobs, \vDobs, \vZobs, \vX)$
\end{enumerate} 
The resulting eight testing methods were assessed under $3\times7\times2$ scenarios defined by the level of predictiveness of covariates on the compliance status (none, medium, high), by seven values of the difference between the outcome mean under control assignment for compliers and never-takers, i.e. $E[Y_i(0)| C_i=1]-E[Y_i(0)| C_i=0]$, and by two values (0 for simulations under $H_0$ and 0.5 for simulations under $H_1$) of a constant complier causal effect, i.e., we set the difference $Y_i(1)-Y_i(0)$ between the outcomes for all compliers.
For each simulation, performance was assessed as the rejection rate at significance level $\alpha$, which corresponds to either power or type \textrm{I} error.
%

\subsection{Data Generating Process}
In all of the simulation sets, we assumed a population of N=500 units posessing a single covariate $X_i\sim\mathcal{N}(0,1)$.
We generated the data and completely randomized $N_T=250$ to the treatment with $Z^{obs}_i=1$ and $N_C=250$ units to  control with $Z^{obs}_i=0$. 
The compliance status follows a probit model conditional on the covariate $X_i$:
\begin{equation}
\label{eq:comp}
C_i=\mathds{1}\{\alpha_0 + \alpha_xX_i+\epsilon_i>0\}   \qquad \epsilon_i\sim\mathcal{N}(0,1) 
\end{equation}
where the coefficient vector $\boldsymbol{\alpha}=[\alpha_0, \alpha_x]$ is varied to result in three different levels of predictiveness: none ($\boldsymbol{\alpha}=[-0.8, 0]$), medium  ($\boldsymbol{\alpha}=[-1.4, 2]$), high  ($\boldsymbol{\alpha}=[-2.8, 5]$). 
The probability of being a complier is then $\Phi(\alpha_0 +\alpha_x X_i)$, where $\Phi(\cdot)$ is the standard normal cumulative function. 
The values of $\alpha_0$ are chosen such that $\pi_c=0.30$, that is, the overall proportion of compliers is always 30\%.
The outcome follows a normal distribution with unitary variance and a mean that depends on the compliance status and the scenario:


\begin{center}
\begin{tabular}{ccc}
$
Y_i(0)\sim
\begin{cases}
\mathcal{N}(\eta_{n}, 1) \quad \text{if $C_i=0$}\\
\mathcal{N}(\eta_{c0}, 1) \!\!\quad \text{if $C_i=1$}
\end{cases}
$&
$
\quad
 Y_i(1)=
\begin{cases}
Y_i(0)\qquad\,\,\,\, \text{if $C_i=0$}\\
Y_i(0)+\tau \quad \text{if $C_i=1$}
\end{cases}
$&
$
\quad
\tau=
\begin{cases}
0 \qquad \text{under $H_0$}\\
0.5 \,\quad \text{under $H_1$}
\end{cases}
$
\end{tabular}
\end{center}

\noindent with $\eta_{n}=0$ and $\eta_{c0}=\{-3,-2,-1,-0.5, \,0, \,0.5, \,1, \,2, \,3\}$.\\
The observed outcome is then given by: $Y_i^{obs}=Y_i(1)Z^{obs}_i+Y_i(0)(1-Z^{obs}_i)$.
In order to have a better control on the overlap between the distributions of the two potential outcomes for never-takers and compliers, we chose to model the outcome as independent from the covariates, even in the two scenarios when these are predictive of the compliance status. 
In scenarios where covariates do affect the outcome, if the individual treatment effect does not depend on covariates then our results should not vary.

\subsection{Bayesian Inference for Imputation of Compliance Statuses}
The computation of posterior predictive $p$-values is performed using an MCMC approach as explained in Section 3. Each iteration comprises an imputation and a permutation step. During the imputation step unknown compliance statuses are drawn from the predictive posterior distribution $\prob(\vC\mid \vYobs, \vDobs, \vZobs)$, with an additional conditioning on $H_0$ and covariates depending on the method used. 
Because the parameters of this distribution are unknown, the predictive posterior distribution of interest has to be averaged over the posterior distribution of the parameters: $\prob(\vC\mid \vYobs, \vDobs, \vZobs)=\int \prob(\vC\mid \vYobs, \vDobs, \vZobs, \vtheta)p(\vtheta\mid \vYobs, \vDobs, \vZobs)d\vtheta$. 
In practice, it is often convenient to make use of a two-stage Gibbs-sampler, which samples the parameters from their full conditional distribution $\prob(\vtheta \mid \vYobs, \vDobs, \vZobs, \vC)$ and then samples the vector of compliance statuses from  $\prob(\vC\mid \vYobs, \vDobs, \vZobs, \vtheta)$. 
This algorithm is known as data augmentation \citep{Tanner:Wong:1987}.
In this simulation study, Bayesian inference is based on correct model specification.
The only misspecification comes from conditioning on the null hypothesis for methods 1 and 3.
Therefore we used the model for the compliance status from \eqref{eq:comp}
and the following model for potential outcomes 
\begin{center}
\begin{tabular}{cc}
$\begin{aligned}
Y_i(0)| C_i=1& \sim\mathcal{N}(\eta_{c1}, \sigma_{c})\\
Y_i(0)| C_i=0 &\sim\mathcal{N}(\eta_{n},  \sigma_{n})
\end{aligned}$
&$\begin{aligned}
Y_i(1)| C_i=1& \sim\mathcal{N}(\eta_{c1},  \sigma_{c})\\
Y_i(1)| C_i=0 &\sim\mathcal{N}(\eta_{n},  \sigma_{n})
\end{aligned}$
\end{tabular}
\end{center}
with means and variances following from exclusion restriction.
At each iteration $k$, after the parameters are drawn from their posterior, the unknown compliance status for units in the control group is imputed as follows:
\begin{equation}
\label{eq:DA}
Pr(C_i=1|Y^{obs}_i,Z_i=0, X_i, \vtheta)=\frac{\phi(Y^{obs}_i; \eta_{c},\sigma_{c} )\Phi(\alpha_0+ \alpha_x X_i)}{\phi(Y^{obs}_i; \eta_{c},\sigma_{c} )\Phi(\alpha_0 + \alpha_x X_i) +\phi(Y^{obs}_i; \eta_{n},\sigma_{n} )(1-\Phi(\alpha_0+ \alpha_x X_i))}
\end{equation}

For imputation methods 1 and 3, we impose the null hypothesis for imputing the compliance status by assuming $\eta_{c0}=\eta_{c1}=\eta_{c}$ in the Bayesian estimation of the parameters. 
Similarly, we set $\alpha_x=0$ for  imputation methods 1 and 2, where we do not take covariates into account. 
We used conjugate prior distributions for the normal model with unknown variance, that is, normal distributions for the means and inverse gamma distributions for the variances:
\[
\eta_{c}, \eta_{c0}, \eta_{c1}, \eta_{n}\sim \mathcal{N}(0,10) \qquad \sigma_{c}, \sigma_{n}\sim \mathcal{IG}(0.1,0.1)
\]

\subsection{Simulation Procedure}
For the eight testing methods defined by the imputation method and the use of test statistic or  discrepancy variable, we, for each of the $3\times7\times2$ scenarios, repeated the following 2000 times:

\begin{enumerate}
\item Generate a randomized sample of $N$ units from the distributions outlined in the generating process (with specific values of $\boldsymbol{\alpha}$ and $\eta_{c0}$ and assuming the null ($\tau=0$) or the alternative hypothesis ($\tau=1$), resulting in $(\vZ^{obs}, \vX, \vC, \vY^{obs})$;
\item Calculate posterior predictive $p$-values using the test statistic or the discrepancy measure via the following steps:
\begin{enumerate}
\item Impute the compliance statuses $\tilde{C}_i$ for units in the control group using one of the four imputation methods. This step is performed by the following two-stage Gibbs-sampling procedure:
\begin{enumerate}
\item Draw a value of the parameters from their posterior distribution, which depends on the imputation method.
\item Use equation \eqref{eq:DA} with the drawn parameters to impute the compliance statuses for units in the control group.
\end{enumerate}
\item Compute the observed test statistic $T(\vYobs, \vD^{obs}, \vZ^{obs})$ or the `observed' discrepancy measure $D(\vYobs, \tilde{\vC}, \vZ^{obs})$.
\item Take a random sample from the set of all possible assignment vectors $\vZ$.
\item Compute the test statistic $T(\vYobs, \tilde{\vC}\vZ, \vZ)$ or discrepancy measure $D(\vYobs, \tilde{\vC}, \vZ)$, based on the imputed compliance statuses and the sampled assignment vector.
\item Compare this test statistic or discrepancy measure with the observed value and output 1 if the new value is larger than the observed and 0 otherwise.
\item Repeat steps (a)-(e) $K=2000$ times and average the outputs of the last 1000 iterations (1000 discarded as burn-in) to get the nominal $p$-values.
\end{enumerate}
\end{enumerate}
Note that we do not need to do a full permutation on the inner loop as the expected value of the single comparison will be that conditional $p$-value.  Averaging these ``estimates'' over the 1000 trials gives our desired overall expectation.

\begin{figure}[ht!]
\begin{center}
\includegraphics[scale=0.6]{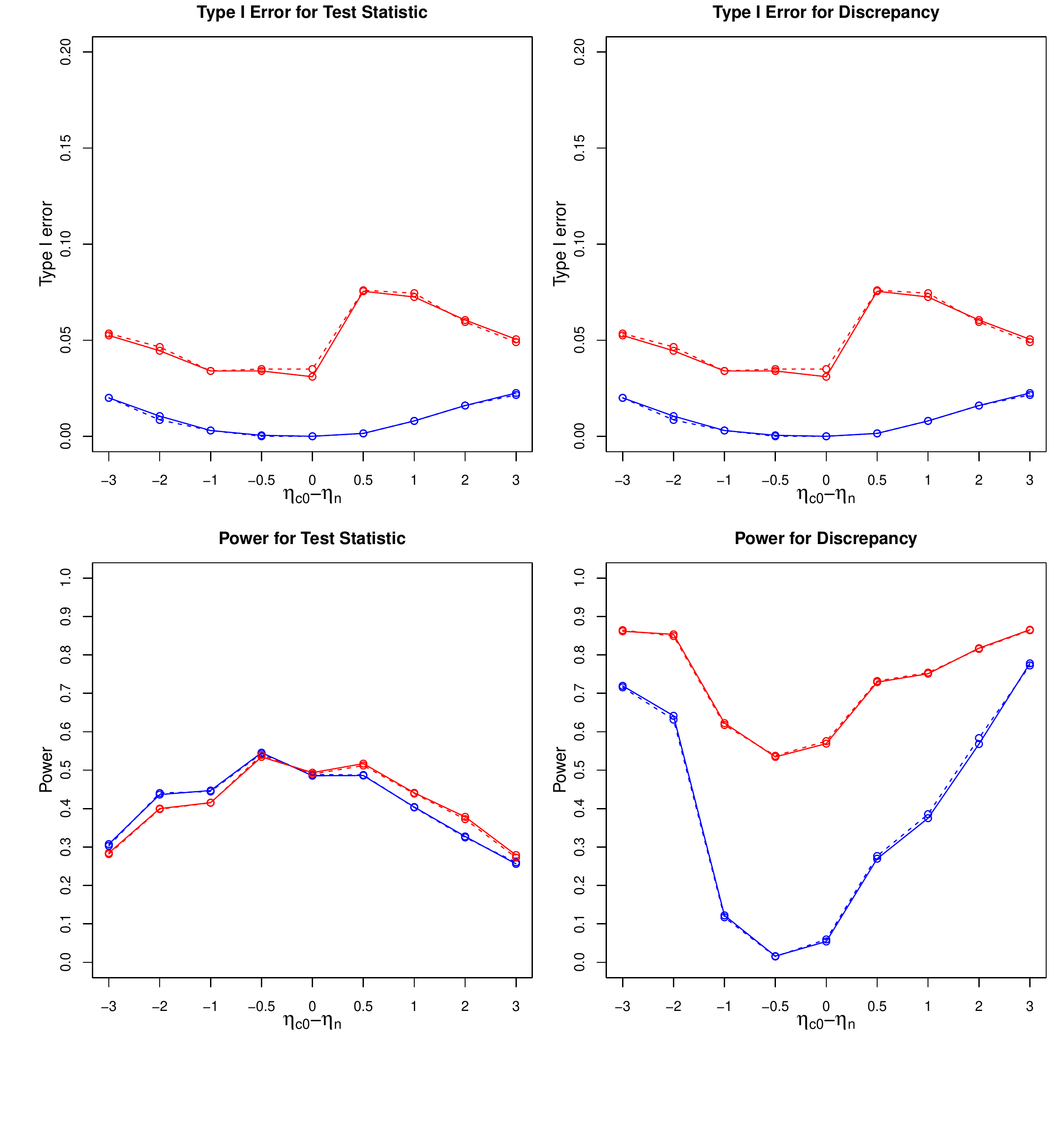}
\vspace{-1.5cm}
\caption{Type I error and power for level of significance $\alpha=0.05$ against $\eta_{c0}-\eta_n$, for both test statistics (left) and discrepancy variables (right) with zero predictiveness of the covariates. The four lines represent the four different methods of imputation of compliance status: $\prob(\vC\mid \vYobs, \vDobs, \vZobs, H_0)$ (solid blue line), $\prob(\vC\mid \vYobs, \vDobs, \vZobs)$ (solid red line),  $\prob(\vC\mid \vYobs, \vDobs, \vZobs, H_0, \vX)$ (dashed blue line), $\prob(\vC\mid \vYobs, \vDobs, \vZobs, \vX)$ (dashed red line).}
\label{fig:zero}
\end{center}
\end{figure}

\section{Results}
\label{sec:res}
To compare the validity and power of the Bayesian FRT for each method, we computed the rejection rates for a test with level $\alpha=0.05$ for simulations conducted under the null hypothesis and the alternative hypothesis, respectively. 
Figure \ref{fig:zero} shows results when the compliance status does not depend on covariates. 
For each imputation method and for both the test statistic and the discrepancy variable, we plot the estimated Type \textrm{I} error and power, based on 2000 simulations, against $\eta_{c0}-\eta_n$, the difference between complier and never-taker means.

Posterior predictive $p$-values based on the test statistic appear to be unaffected by the imputation methods. 
In principle the test statistic depends on the imputed compliance statuses through the corresponding proportion of compliers $\hat{\pi}_c$. 
Nevertheless, this quantity is robust to model misspecification, given that, thanks to randomization, $Pr(C_i=1)=Pr(D_i=1|Z_i=1)$.
Therefore, Bayesian tests based on this test statistic appear to have a size around the nominal level for any imputation method and for any distance between compliers and never-takers in the control group. 
In fact, histograms of the distributions of such $p$-values (not shown) suggest a uniform distribution. 
The power of these tests, however, decreases as $\eta_{c0}-\eta_n$ gets larger. 
This occurs because the variance of $\widehat{ITT}_Y$ depends on the difference in outcomes between compliers and never-takers.

Unlike test statistic $p$-values, discrepancy-based $p$-values are very sensitive to the compliance imputation method, given that at each iteration the discrepancy measure assumes the imputed compliance status for every unit as the true status. 
\begin{figure}[ht!]
\begin{center}
\includegraphics[scale=0.6]{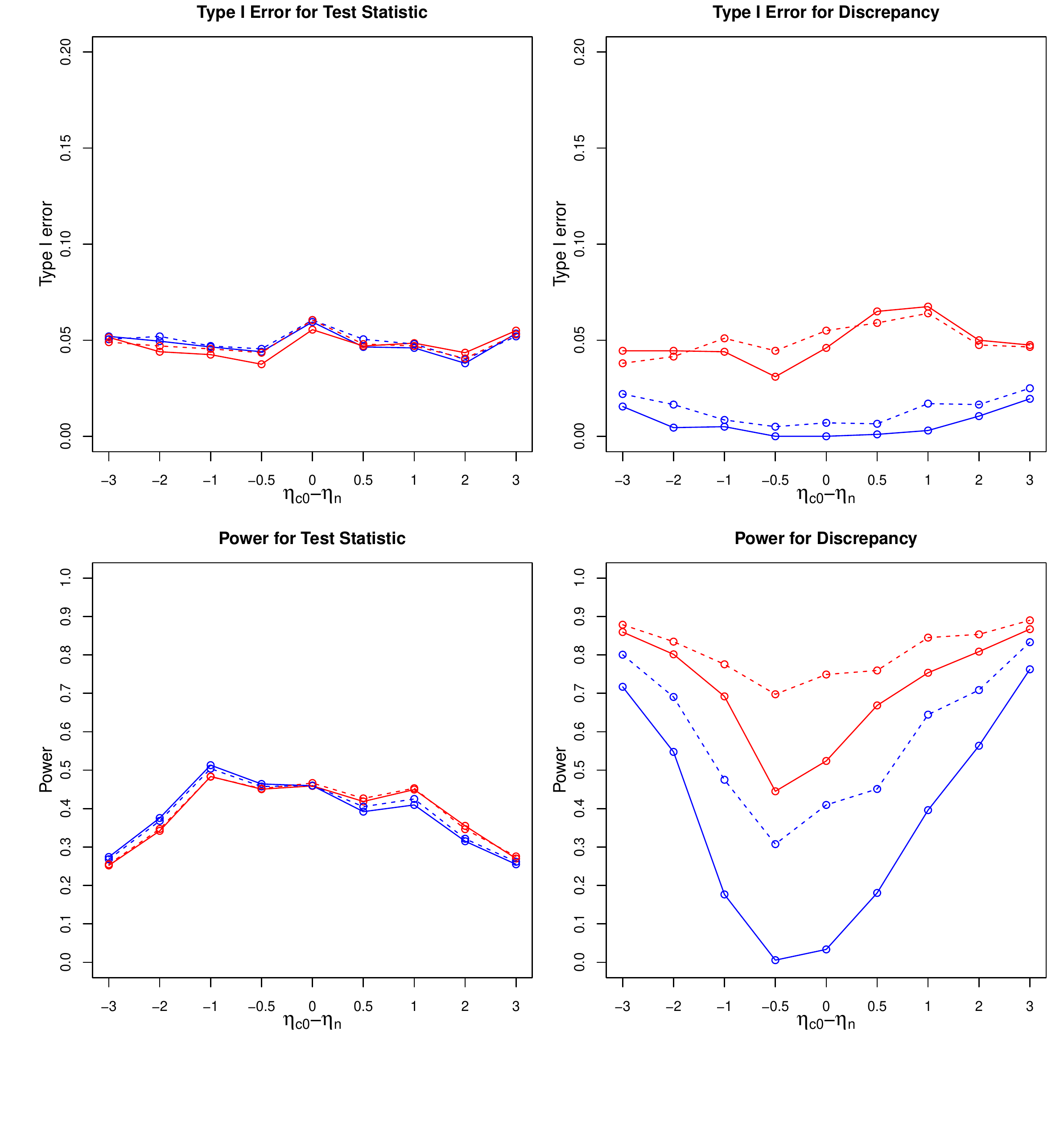}
\vspace{-1.5cm}
\caption{Type I error and power for level of significance $\alpha=0.05$ against $\eta_{c0}-\eta_n$, for both test statistics (left) and discrepancy variables (right) with medium predictiveness of the covariates. The four lines represent the four different methods of imputation of compliance status: $\prob(\vC\mid \vYobs, \vDobs, \vZobs, H_0)$ (solid blue line), $\prob(\vC\mid \vYobs, \vDobs, \vZobs)$ (solid red line),  $\prob(\vC\mid \vYobs, \vDobs, \vZobs, H_0, \vX)$ (dashed blue line), $\prob(\vC\mid \vYobs, \vDobs, \vZobs, \vX)$ (dashed red line).}
\label{fig:med}
\end{center}
\end{figure}
As expected, when imputing conditional on the null hypothesis, the test is conservative with a Type \textrm{I} error around 0.01. 
Also as hypothesized, when outcomes for compliers and never-takers in the control group are close, the use of discrepancy variables fails dramatically to reject a false null hypothesis under the alternative. 
The difficulty in disentangling the mixture between never-takers and compliers when these have similar outcomes, in combination with the exclusion restriction assumption (implying $E\{Y_i(0)|C_i=0\}=E\{Y_i(1)|C_i=0\}$) and the null hypothesis (implying $E\{Y_i(0)|C_i=1\}=E\{Y_i(1)|C_i=1\}$), leads to an overestimation of the mean $\eta_{c0}$ for control compliers and, as a consequence, units in the control group imputed as compliers tend to have outcomes that are closer to those for treated compliers. 
In particular, when $\eta_{c0}-\eta_{n}=0$ or $-0.5$ the power of the test goes to zero.

Imputing with $\eta_{c0}$ and $\eta_{c1}$ unconstrained, i.e., imputing without conditioning on the null, partially corrects this phenomenon. 
Under this alternate imputation strategy, the discrepancy-based FRT-PP yields greater power than when imputing under the null, with a minimum of $51\%$ when $\eta_{c0}-\eta_{n}=0$. 
Moreover, the FRT-PP with an unconstrained imputation method also outperforms test statistic-based $p$-values with an average gain in power of $30\%$.
Unfortunately, this gain in power comes with a substantial price: there is a range of values for $\eta_{c0}-\eta_{n}$ where the test is invalid.
Other simulations (not shown) reveal that this loss in validity is even higher when the proportion of compliers is lower. 


The above is without any predictive covariate.
We next examine whether improved prediction of compliance status changes these patterns.
Figure \ref{fig:med} shows results for the scenario where covariates predict compliance status with a medium level of predictiveness. 
Performance of statistic-based tests is similar to the previous case. 
Discrepancy variables, however, are more robust: the problem of invalid tests when imputing the compliance statuses without imposing the null is attenuated. 
Furthermore, when covariates are used as predictors of the compliance status in the imputation step, power can greatly increase for all discrepancy-based methods. 
These improvements in validity and power are even more visible when the level of predictiveness of the covariates is higher, as shown in Figure \ref{fig:high}.

\begin{figure}[ht!]
\begin{center}
\includegraphics[scale=0.6]{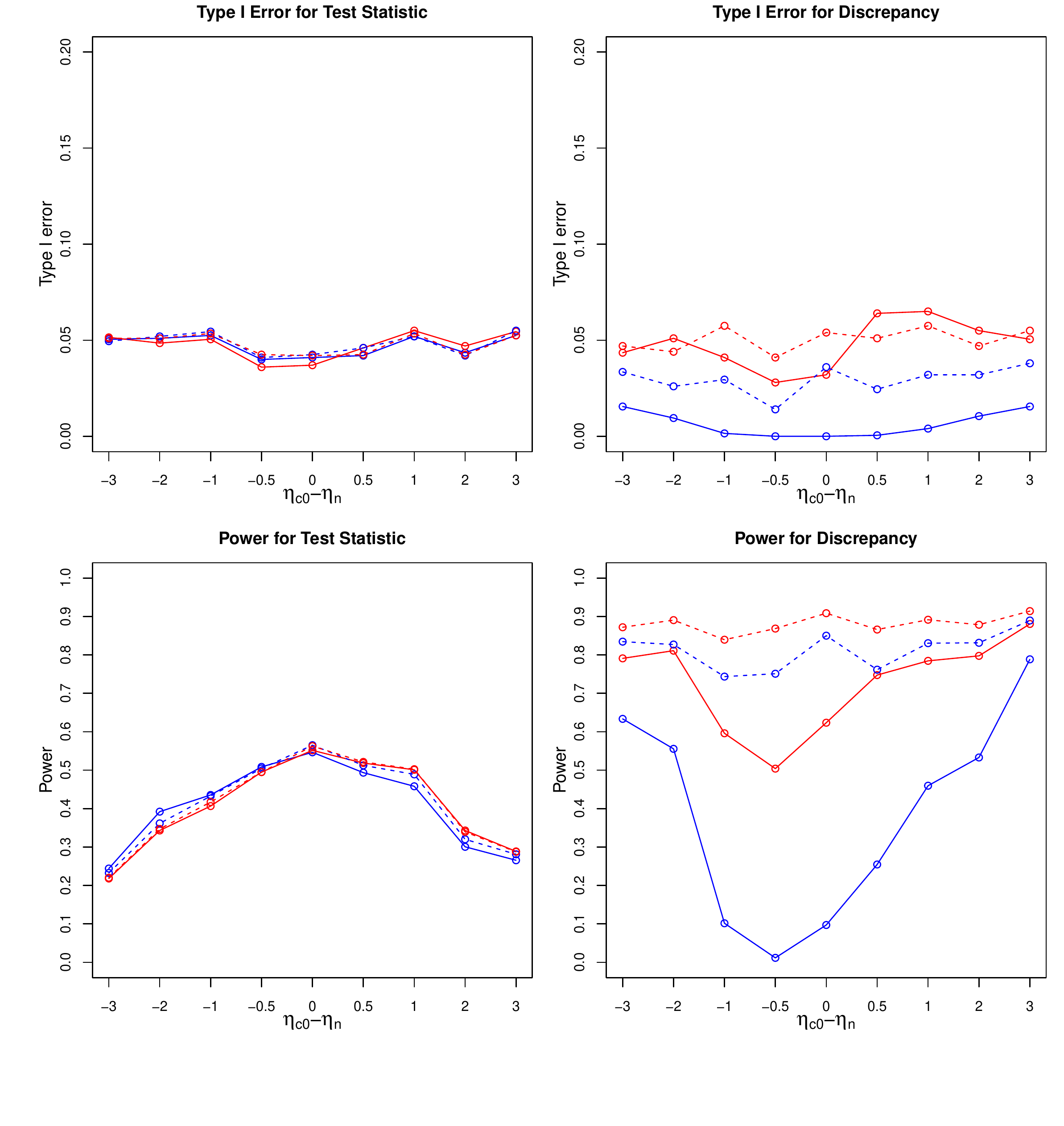}
\vspace{-1.5cm}
\caption{Type I error and power for level of significance $\alpha=0.05$ against $\eta_{c0}-\eta_n$, for both test statistics (left) and discrepancy variables (right) with high predictiveness of the covariates. The four lines represent the four different methods of imputation of compliance status: $\prob(\vC\mid \vYobs, \vDobs, \vZobs, H_0)$ (solid blue line), $\prob(\vC\mid \vYobs, \vDobs, \vZobs)$ (solid red line),  $\prob(\vC\mid \vYobs, \vDobs, \vZobs, H_0, \vX)$ (dashed blue line), $\prob(\vC\mid \vYobs, \vDobs, \vZobs, \vX)$ (dashed red line).}
\label{fig:high}
\end{center}
\end{figure}

%

\section{Comparison with Model-based Tests}
\label{sec:bb}
As we have seen, the FRT-PP methods based on discrepancies heavily rely on the imputation model. 
The advantage of Fisher randomization tests as model-free tests is then somewhat lost. 
Averaging Fisher randomization $p$-values over the predictive posterior distribution of the compliance status 
depends on a posterior predictive distribution that, in turn, depends on the posterior distribution of the model parameters. 
Given this, one might think of just relying on the posterior distribution of the parameters, which in this case would be $\eta_{c1}-\eta_{c0}$. 
Model-based $p$-values for testing a zero treatment effect would then be $Pr\{\eta_{c1}-\eta_{c0}\geq 0 |  \vYobs, \vDobs, \vZobs, \vX\}$. 
Using these Bayesian model-based $p$-values to test the null hypothesis is arguably model-dependent, but so is the FRT-PP based on discrepancies.
The direct Bayesian approach thus seems superior as it does not require a permutation step making it less computationally demanding as well as more transparent. 

\begin{figure}[ht!]
\begin{center}
\includegraphics[scale=0.6]{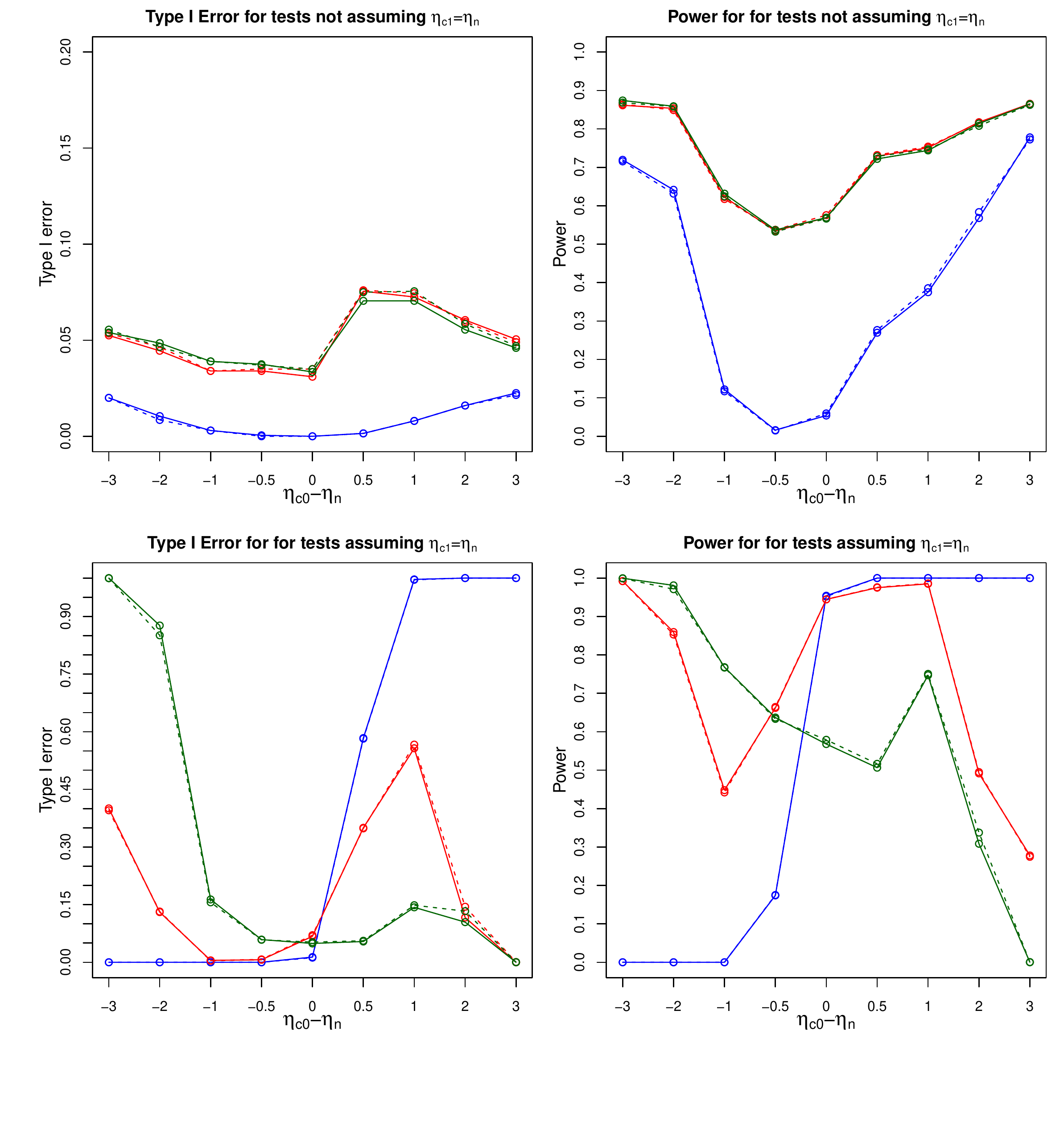}
\vspace{-1.5cm}
\caption{Type I error and power for level of significance $\alpha=0.05$ against $\eta_{c0}-\eta_n$, for FRT-PP based on discrepancy variables and model-based tests with low predictiveness of the covariates. The six lines represent FRT-PP with the four different methods of imputation of compliance status, $\prob(\vC\mid \vYobs, \vDobs, \vZobs, H_0)$ (solid blue line), $\prob(\vC\mid \vYobs, \vDobs, \vZobs)$ (solid red line),  $\prob(\vC\mid \vYobs, \vDobs, \vZobs, H_0, \vX)$ (dashed blue line), $\prob(\vC\mid \vYobs, \vDobs, \vZobs, \vX)$ (dashed red line), and model-based tests not including and including covariates in the compliance model (solid green and dashed green lines, respectively).} 
\label{fig:Bayes_zero}
\end{center}
\end{figure}

On the other hand, such a model-based test could be relying more heavily on model assumptions than the FRT-PP approach, given that it strictly depends on good estimation of both $\eta_{c1}$ and $\eta_{c0}$. 
By contrast, in one-side noncompliance FRT-PP does not directly depend on the estimation $\eta_{c1}$, and only uses $\eta_{c0}$ and $\eta_n$ indirectly to impute compliance statuses in the control group. 
Consequently, we might expect model misspecification to affect model-based $p$-values more severely.  
We explore this under an extreme case of misspecification, that is, when we impose the assumption that the outcome mean under treatment is the same for compliers and never-takers in the estimation step:
\begin{equation}
\label{ass: eq}
E[Y_i(1)|C_i=0]=E[Y_i(1)|C_i=1]
\end{equation}
This assumption might be plausible in different situations. 
An example is when never-takers do not take treatment due to access to some other, similar, treatment that we do not see.
Imposing this assumption when it is not true will compromise the estimation of $\eta_{c1}$ and $\eta_{n}$, which will indirectly impact the estimation of $\eta_{c0}$. 

Figure \ref{fig:Bayes_zero} compares the Type \textrm{I} error and power of discrepancy-based FRT-PP tests and model-based tests under this misspecification when compliance status is not predicted by covariates. 
First, when we have correct model specification, model-based tests have the same performance as discrepancy-based FRT-PP using an unconstrained imputation method, as illustrated by the top row of Figure~\ref{fig:Bayes_zero}.
Note the overlapping lines corresponding to the model-based approach and the unconstrained imputation approach with and without covariates (imposing the null results in a reduction of rejection rates across the board).

Under the misspecification induced by Assumption~\ref{ass: eq}, however, performance patterns diverge.  
This is the bottom row of Figure~\ref{fig:Bayes_zero}.
The validity of model-based tests is compromised when $\eta_{c0}<\eta_{n}$ because $\eta_{c1}$, which should be equal to $\eta_{c0}$ under the null, is instead assumed to be equal to $\eta_{n}$. 
When $\eta_{c0}-\eta_{n}=-3$ false rejection rate gets as high as 100\% for the model-based method. 
Discrepancy-based tests also do not fare well, however.
For the methods with imputation under the null, if $\eta_{c0}-\eta_{n} < 0$ validity is preserved but power is  wiped out .
As $\eta_{c0}-\eta_{n}$ grows increasingly positive, rejection rates jump regardless of whether the null is true or not.
Overall, imposing assumption \ref{ass: eq}, together with the exclusion restriction for never-takers and the null hypothesis, leads to a complete failure of the discrepancy-based FRT-PP that impute compliance statuses in the control group conditioning on the null. 

When imputing without imposing the null, the pattern is somewhat irregular.
For $\eta_{c0}-\eta_{n}<0$ under the null, the size of the tests becomes conservative for a low difference between compliers and never-takers and then increases as the difference gets higher. 
This is due to an underestimation of  $\eta_{c0}$ when $\eta_{c0}-\eta_{n}<<0$. Conversely, when $\eta_{c0}-\eta_{n}>0$, the size of the tests increases for a low difference and decreases as the difference gets higher. 
The size of these tests, while under the nominal level when $\eta_{c0}-\eta_{n}\leq 0$, jumps immediately out of the range of validity when the outcome mean for compliers is greater than the one for never-takers.
None of the tests perform well under this misspecification.  
When comparing the left and right sides, the pattern of rejection rates appears to be more about the value of $\eta_{c0}-\eta_{n}$ than the actual hypothesis being tested.


\begin{figure}[ht!]
\begin{center}
\includegraphics[scale=0.6]{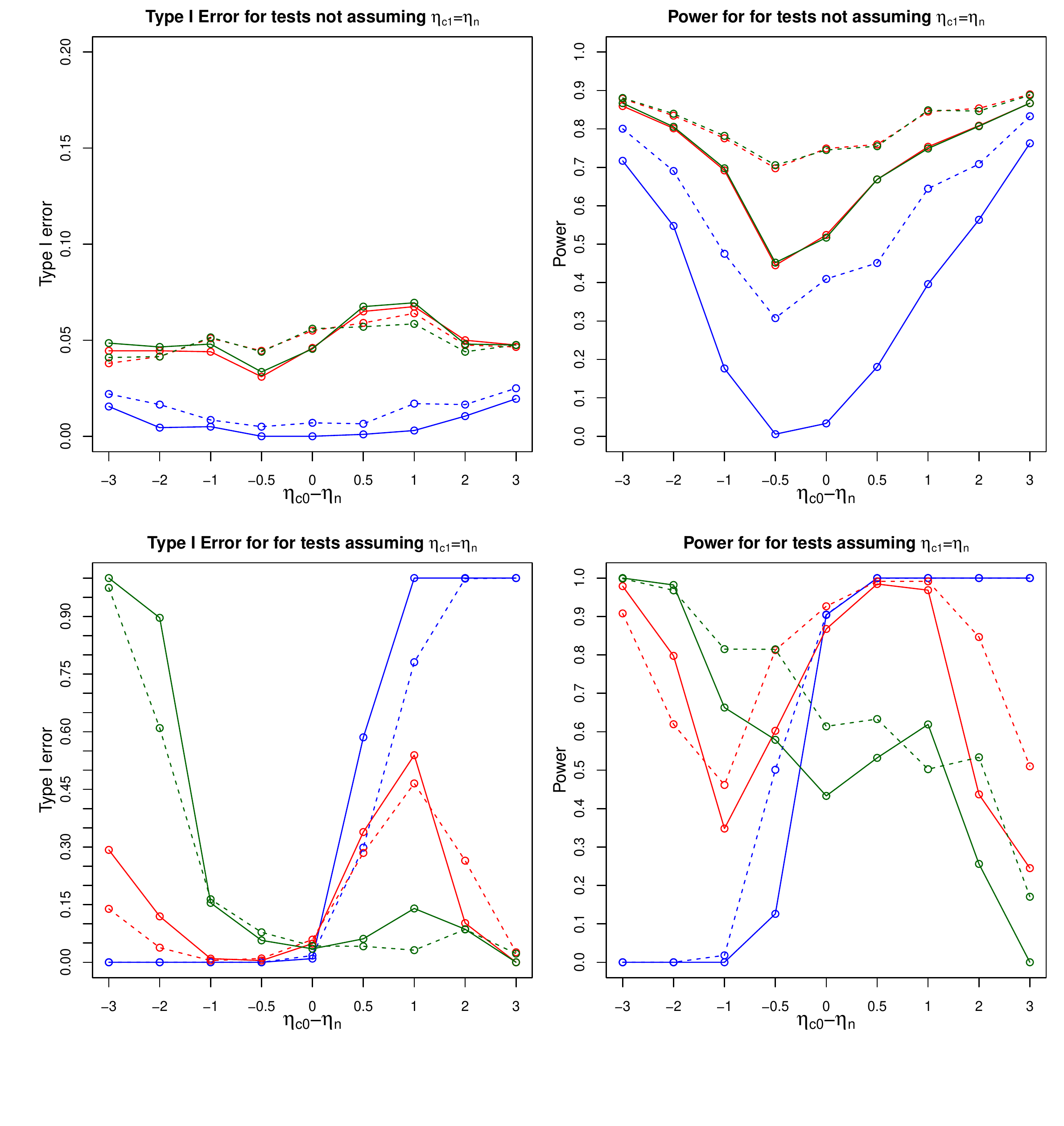}
\vspace{-1.5cm}
\caption{Type I error and power for level of significance $\alpha=0.05$ against $\eta_{c0}-\eta_n$, for FRT-PP based on discrepancy variables and model-based tests with medium predictiveness of the covariates. The six lines represent FRT-PP with the four different methods of imputation of compliance status, $\prob(\vC\mid \vYobs, \vDobs, \vZobs, H_0)$ (solid blue line), $\prob(\vC\mid \vYobs, \vDobs, \vZobs)$ (solid red line),  $\prob(\vC\mid \vYobs, \vDobs, \vZobs, H_0, \vX)$ (dashed blue line), $\prob(\vC\mid \vYobs, \vDobs, \vZobs, \vX)$ (dashed red line), and model-based tests not including and including covariates in the compliance model (solid green and dashed green lines, respectively).} 
\label{fig:Bayes_med}
\end{center}
\end{figure}


Figure \ref{fig:Bayes_med} and Figure \ref{fig:Bayes_high} compares the model-based and FRT-PP approaches under medium and high predictiveness, respectively.
Under correct specification (i.e., when \ref{ass: eq} is not imposed) model-based tests still have the same performance as discrepancy-based FRT-PP using an unconstrained imputation method, with and without the use of covariates.
When model parameters are estimated under assumption \ref{ass: eq}, model-based and discrepancy-based FRT-PP are affected as previously, when covariates are not included in the compliance model. 
The use of compliance-predictive covariates improves validity and reduces the power loss for all tests.
Nevertheless, even with a high level of predictiveness, both model-based tests and discrepancy-based FRT-PP imposing the null still have very low performance for wide ranges of values.
The use of covariates does not always ensure validity for discrepancy-based FRT-PP that do not impose the null. 
In fact, when $\eta_{c0}-\eta_{n}\in\{0.5,2\}$ the size of these tests is still higher than 15\%, well above the nominal $\alpha=0.05$. 
That being said, these results also do suggest that unconstrained discrepancy-based FRT-PP can be more robust to model misspecification than both model-based tests and discrepancy-based FRT-PP conditioning on the null.
In particular, they do not suffer the of complete invalidity as the other approaches (peaking at 60\% Type I error vs. the others achieving near 100\% error rates), while maintaining a generally higher level of power under the alternative.

\begin{figure}[ht!]
\begin{center}
\includegraphics[scale=0.6]{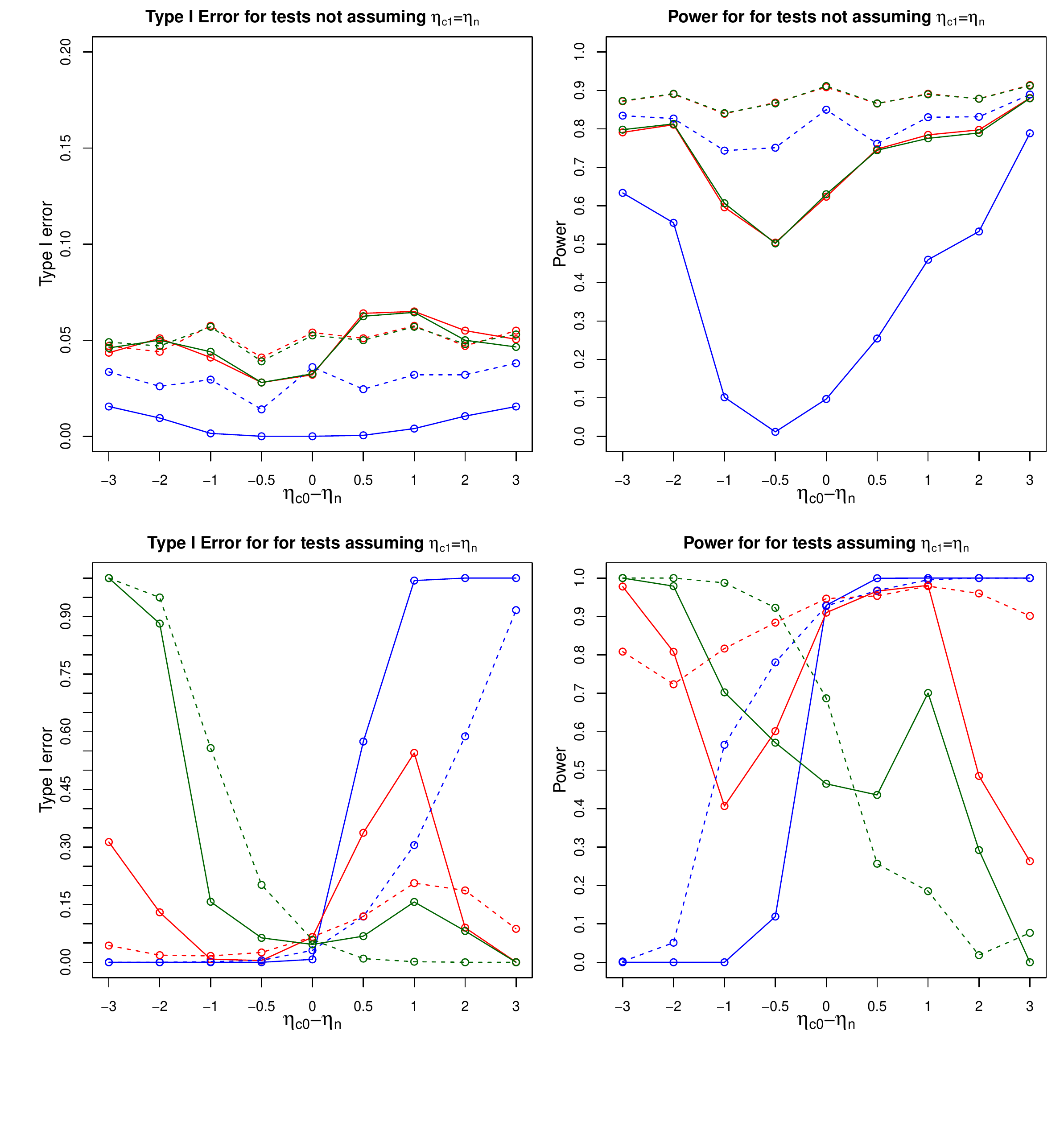}
\vspace{-1.5cm}
\caption{Type I error and power for level of significance $\alpha=0.05$ against $\eta_{c0}-\eta_n$, for FRT-PP based on discrepancy variables and model-based tests with high predictiveness of the covariates. The six lines represent FRT-PP with the four different methods of imputation of compliance status, $\prob(\vC\mid \vYobs, \vDobs, \vZobs, H_0)$ (solid blue line), $\prob(\vC\mid \vYobs, \vDobs, \vZobs)$ (solid red line),  $\prob(\vC\mid \vYobs, \vDobs, \vZobs, H_0, \vX)$ (dashed blue line), $\prob(\vC\mid \vYobs, \vDobs, \vZobs, \vX)$ (dashed red line), and model-based tests not including and including covariates in the compliance model (solid green and dashed green lines, respectively).} 
\label{fig:Bayes_high}
\end{center}
\end{figure}

\section{Concluding Remarks}
\label{sec:conc}
Fisher randomization tests (FRTs) are concerned with testing hypothesis regarding the effect of treatments. By exploiting the known assignment mechanism, they work simply by enumerating all possible values of a specified test statistic under the null hypothesis and comparing this distribution to the observed value.
In noncompliance settings, one can impute the unknown compliance statuses with a Bayesian framework and conduct randomization tests with the complete data. The resulting FRT-PP then averages over the posterior distribution of the unknown compliance statuses. 
Here, we compared the use of discrepancy measures to classical test statistics within this framework.
Being parameter-dependent, discrepancies more heavily depend on the imputation model, including the imposition of the null hypothesis. 
We investigated the resulting trade-off between the size and the power of the test. 

Discrepancy-based FRT-PP that use an imputation method under the null are fairly conservative, and are generally valid testing procedures.
Unfortunately, however, we found situations where misclassification of compliers and never-takers, due to imposing the null, led to a severe reduction of power. 
This problem can be overcome by using an unconstrained imputation method, which on average achieves higher power than both discrepancy-based FRT-PP under the null and statistic-based FRT-PP. 
Unfortunately, this unconstrained imputation can result in invalid tests, even in scenarios where the model is correctly specified.

With regard to the use of test statistics, the size of the test stayed close to the nominal levels for all scenarios. 
Power decreases as the outcomes for compliers in the control group move away from those for never-takers, but is, on average, greater than the one given by discrepancy-based FRT-PP assuming the null. 
Overall, our test statistic, compared to our discrepancy measure, seems less affected by the imputation method.  The test statistic, in fact, depends on the imputation model only through the estimated proportion of compliers. As a natural consequence, statistic-based tests are expected to be more robust to model misspecification as long as the proportion of imputed compliers is not significantly compromised.  

Discrepancies depend tremendously on the imputation method and can be largely affected by model misspecification. 
Therefore, as a recommendation for practitioners, if the model is likely misspecified (e.g., important covariates are missing, measures of goodness of fit are not satisfactory, etc.) then use FRT-PP based on test statistics, not discrepancy measures, to maintain validity.
That being said, if we are confident in our posited model, the use of discrepancy-based tests could in principle improve power. 
This is not always the case if the imputation is performed under the null, as this method can oftentimes lead to a large loss of power. 
For the purpose of power gain, imputation methods that do not impose the null hypothesis seem to be more appropriate, save for the serious concern of an elevated risk of validity.
The use of predictive covariates, if present, does attenuate this risk as well as it does increase power for either imputation method.
Overall, with a high or medium level of predictiveness, we recommend leaving the imputation method unconstrained.  

As a final investigation, we compared the discrepancy-based FRT-PP method to full Bayesian model-based testing.
We found that a model-based test was comparable to a discrepancy-based test that does not impose the null hypothesis.
However, the latter FRT-PP method was more robust to misspecification, giving evidence that the additional permutation step can somewhat protect a researcher from model specification issues when doing inference.
These simulations are for a specific class of data generation processes, and further work needs to be done to more fully understand this novel form of statistical inference.
Certainly there are substantial dangers in taking these approaches, but there is also evidence that there is a lot to be gained.
\cleardoublepage
\newpage

\renewcommand{\refname}{\scshape \textbf{References}}

\bibliographystyle{jasa}

\begin{thebibliography}{7} \setlength{\itemsep}{0.3em}
\expandafter\ifx\csname natexlab\endcsname\relax\def\natexlab#1{#1}\fi
\bibitem[{Angrist et~al.(1996)}]{Angrist:1996}
\textsc{Angrist, J.~D.}, \textsc{Imbens,G.~W.}, \& \textsc{Rubin, D.~B.} (1996). \newblock{Identification of causal effects using instrumental
variables (with discussion)}. \newblock \textit{Journal of the American Statistical Association}, 91, 444--472.

\bibitem[{Berger \& Boos(1994)}]{Berger:Boos:1994}
\textsc{Berger, R.~L.} \& \textsc{Boos, D.~D.} (1994). \newblock{P values maximized over a confidence set for the nuisance parameter}. \newblock \textit{Journal of the American Statistical Association}, 89, 1012--1016.

\bibitem[{Ding et~al.(2015)}]{Ding:2015}
\textsc{Ding, P., Feller, A., and Miratrix, L.} (2015). \newblock{Randomization Inference for Treatment Effect Variation}. \newblock 

\bibitem[{Fisher(1925)}]{Fisher:1925}
\textsc{Fisher, R.A.} (1925). \newblock{Statistical Methods for Research Workers}. \newblock \textit{1st
ed. Oliver and Boyd, Edinburgh}.

\bibitem[{Fisher(1926)}]{Fisher:1926}
\textsc{Fisher, R.A.} (1925). \newblock{The arrangement of field experiments}. \newblock \textit{J.Ministry of Agriculture of Great Britain}, 33, 503--513.

\bibitem[{Fisher(1935)}]{Fisher:1935}
\textsc{Fisher, R.A.} (1925). \newblock{The design of experiments}. \newblock \textit{Edinburgh: Oliver and Boyd.}.

\bibitem[{Frangakis \& Rubin(2002)}]{Frangakis:Rubin:2002}
\textsc{Frangakis, C.~E.} \& \textsc{Rubin, D.~B.} (2002). \newblock{Principal stratification in causal inference}. \newblock  \textit{Biometrics}, 58,
21--29.

\bibitem[{Gelman et~al.(1996)}]{Gelman:1996}
\textsc{Gelman, A., Meng, X. L., and Stern, H. S.} (1996). \newblock{Posterior predictive assessment
of model fitness via realized discrepancies (with discussion)}. \newblock \textit{Statistica Sinica}, 6, 733--807.


\bibitem[{Gelman(2013)}]{Gelman:2013}
\textsc{Gelman, A.} (2013). \newblock{Two simple examples for understanding
posterior $p$-values whose distributions are far from uniform}. \newblock \textit{Electronic Journal of Statistics}, 7, 2595--2602.

\bibitem[{Guttman(1967)}]{Guttman:1967}
\textsc{Guttman, I.} (1967). \newblock{The Use of the Concept of a Future Observation in Goodness-of-Fit Problems}. \newblock \textit{Journal of the Royal Statistical Society} B, 29(1) 83--100.

\bibitem[{Imbens \& Angrist(1994)}]{Imbens:Angrist:1994}
\textsc{Imbens, G.~W.} \& \textsc{Angrist, J.~D.} (1994).  \newblock{Identification and estimation of local average treatment effects}. \newblock \textit{Econometrica}, 62, 467--476.

\bibitem[{Imbens \& Rubin(1997)}]{Imbens:Rubin:1997}
\textsc{Imbens, G.~W.} \& \textsc{Rubin, D.~B.} (1997). \newblock{Bayesian inference for causal effects in randomized experiments with noncompliance}. \newblock \textit{Annals of Statistics}, 25, 305--327.

\bibitem[{Holland(1986)}]{Holland:1986}
\textsc{Holland, P.~W.} (1986).  \newblock{Statistics and Causal Inference}. \newblock \textit{Journal
of the American Statistical Association}, 81(396), 945--960.

\bibitem[{Meng(2000)}]{Meng:2000}
\textsc{Bayarri, M. J. and Berger, J.} (2000). \newblock{$p$-values for composite null models.}. \newblock \textit{Journal
of the American Statistical Association}, 95, 1127--1142..

\bibitem[{Meng(1994)}]{Meng:1994}
\textsc{Meng, X.L.} (1994). \newblock{Posterior predictive $p$-values}. \newblock \textit{Annals of Statistics}, 22, 1142--1160.

\bibitem[{Neyman(1923)}]{Neyman:1923}
\textsc{Neyman, J.} (1923). \newblock{On the Application of Probability Theory to Agricultural
Experiments. Essay on Principles. Section 9.}  \newblock \textit{Roczniki Nauk Rolniczych
Tom X [in Polish]; translated in Statistical Science}, 5, 465--480. 

\bibitem[{Neyman(1934)}]{Neyman:1934}
\textsc{Neyman, J.} (1934). \newblock{On two different aspects of the representative method: The method of stratified sampling and the method of purposive selection with discussion.}  \newblock \textit{Journal of the Royal Statistical Society}, 97, 558--625. 

\bibitem[{Nolen \& Hudgens(2011)}]{Nolen:Hudgens:2011}
\textsc{Nolen, T.} \& \textsc{Hudgens, M.~G.} (2011). \newblock{Randomization-based inference within principal strata.}. \newblock \textit{Journal of the American Statistical Association}, 106, 581--593.

\bibitem[{Robins et~al.(2000)}]{Robins:2000}
\textsc{Robins, J. M., Vaart, A., and Ventura, V. } (2000). \newblock{Asymptotic distribution of p
values in composite null models}. \newblock \textit{Journal of the American Statistical Association}, 95, 1143--1156.

\bibitem[{Rubin(1974)}]{Rubin:1974}
\textsc{Rubin, B.~D.} (1974).  \newblock{Estimating causal effects of treatments in randomized and non randomized studies}. \newblock \textit{Journal of Educational Psychology} 66, 688--701.

\bibitem[{Rubin(1978)}]{Rubin:1978}
\textsc{Rubin, B.~D.} (1978a).  \newblock{Bayesian inference for causal effects}. \newblock \textit{Annals of Statistics}, 6, 34--58.

\bibitem[{Rubin(1981)}]{Rubin:1981}
\textsc{Rubin, D.B.} (1981). \newblock{Estimation in Parallel Randomized Experiments}. \newblock \textit{Journal of Educational Statistics}, 6(4), 377--401.

\bibitem[{Rubin(1984)}]{Rubin:1984}
\textsc{Rubin, D.B.} (1984). \newblock{Bayesianly Justifiable and Relevant Frequency Calculations for the Applied Statistician}. \newblock \textit{Annals of Statistics}, 12(4), 1151--1172.

\bibitem[{Rubin(1996a)}]{Rubin:1996a}
\textsc{Rubin, D.B.} (1996a). \newblock{Discussion of ``Posterior predictive $p$-values?'' by Gelman, A., Meng, X. L. and Stern, H.}. \newblock \textit{Statistica Sinica}, 6, 787--792.

\bibitem[{Rubin(1996b)}]{Rubin:1996b}
\textsc{Rubin, D.B.} (1996b). \newblock{Multiple imputation after 18+ years (with discussion)}. \newblock \textit{Journal of the American Statistical Association}, 91, 473--520. 

\bibitem[{Rubin(1998)}]{Rubin:1998}
\textsc{Rubin, D.B.} (1998). \newblock{More powerful randomization-based $p$-values in double-blind trials with non-compliance}. \newblock \textit{Statistics in Medicine}, 17(3), 371--85.


\bibitem[{Tanner \& Wong(1987)}]{Tanner:Wong:1987}
\textsc{Tanner, M.~A.} \& \textsc{Wong, W.~H.}(1987). \newblock{The Calculation of Posterior Distributions by Data Augmentation (with discussions)}. \newblock\textit{Journal of the American Statistical Association}, 82, 528--550.


\end{thebibliography}

\end{document}